# The Performance and Long Term Stability of the D0 Run II Forward Muon Scintillation Counters   FERMILAB-PUB-14-128-E

V. Bezzubov [1], D. Denisov[2], V. Evdokimov[1], V. Lipaev[1], A. Shchukin[1], I. Vasilyev[1]
[1]Institute for High Energy Physics, Protvino, Russia
[2]Fermi National Accelerator Laboratory, Batavia IL, USA

## Abstract

The performance of the D0 experiment forward muon scintillation counters system during Run II of the Tevatron from 2001 to 2011 is described. The system consists of 4214 scintillation counters in six layers. The long term stability of the counters amplitude response determined using LED calibration system and muons produced in proton–antiproton collisions is presented. The averaged signal amplitude for counters of all layers has gradually decreased over ten years by 11%. The reference timing, determined using LED calibration, was stable within 0.26 ns. Averaged value of muon timing peak position was used for periodic D0 clock signal adjustments to compensate seasonal drift caused by temperature variations. Counters occupancy for different triggers in physics data collection runs and for minimum bias triggers are presented. The single muon yields versus time and the luminosity dependence of yields were stable for the forward muon system within 1% over 10 years.

Keywords: D0 experiment, scintillation counters, muon system, monitoring

## 1. Introduction

The long term stability of scintillation counters and other detectors is important for large collider detectors designed for years and even decades of operation. There is possibility of significant performance degradation of scintillation counters with time and/or absorbed radiation [1,2]. Results of performance and stability monitoring of the D0 experiment forward muon scintillation counters and front-end electronic during Tevatron Run II are described in this article, including stability of amplitude and timing response, reasons for counters failures, occupancy of the counters during collider operation as well as use of muons produced in proton–antiproton collisions to monitor long term stability of the muon detector and triggers.

The D0 forward muon system covers pseudorapidity $\eta$ range $1.0<|\eta|<2.0$ and consists of trigger scintillation (pixel) counters system and the forward muon tracking system of mini-drift tubes (MDT) as described in references [3,4] and shown in Fig.1. The "pixel" counter system resides beyond the calorimeter and consists of 4214 trapezoidal-shaped scintillation counters in 6 layers, North A and South A layers in front of 1.8 T iron toroids, each followed by similar layers B and C after the toroids. Muons originating from the interaction region traverse the three counter layers, A, B, and C, for North or South side around the interaction point. Counters in layers are grouped into eight octants, each octant containing up to 96 counters. The A-layers and the B-layers are mounted on the inside and outside faces of the two end toroids respectively, while the C-layers are mounted on two separate steel structures mounted on sidewalks of the detector hall. The C-layer, the largest among the three layers, is approximately 12 x 10 m$^2$. A photograph of C-layer pixel counters is presented in Fig. 2.

In A and B-layers, counters are placed in 12 concentric zones in the $\eta$ direction, C-layers consists of 11 zones because of limited space. The $\eta$-segmentation is 0.12 for the nine inner zones and 0.07 for the three outer zones. The $\varphi$-segmentation is 4.5°. The size of the smallest counters in the A-layer is 9 x 14 cm$^2$ and the size of the largest counters in the C-layer is 60 x 110 cm$^2$. The counters use 12.8-mm-thick BICRON 404A [5] scintillator plates of trapezoidal shape and WLS bars for light collection. WLS bars material is SOFZ-105 [6] and based on PMMA (polymethylmethacrylate) plastic containing wavelength shifting fluorescent dopant Kumarin 30.

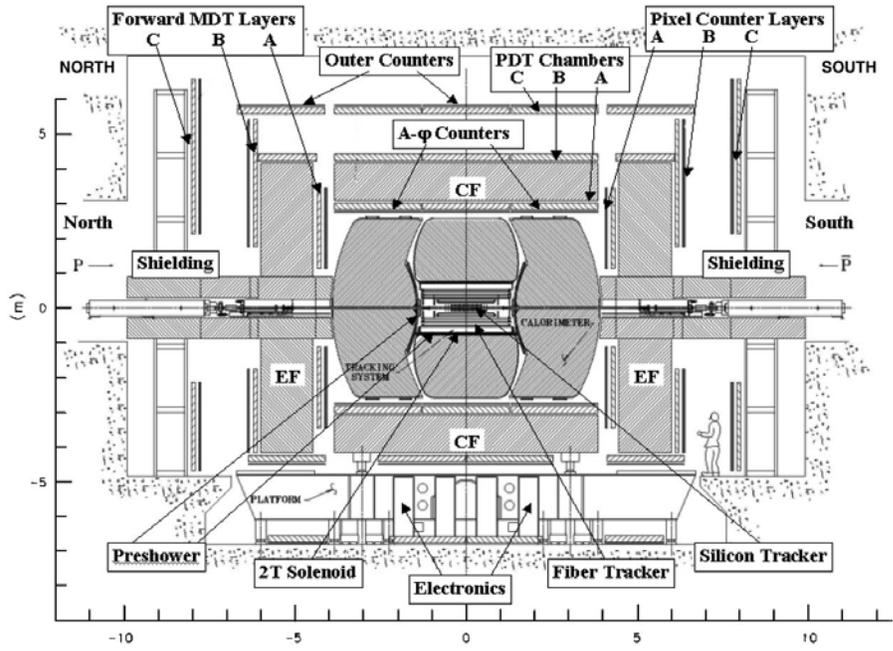

Fig. 1. Cross-sectional view of the D0 Run II detector. The main components of the D0 muon system are identified.

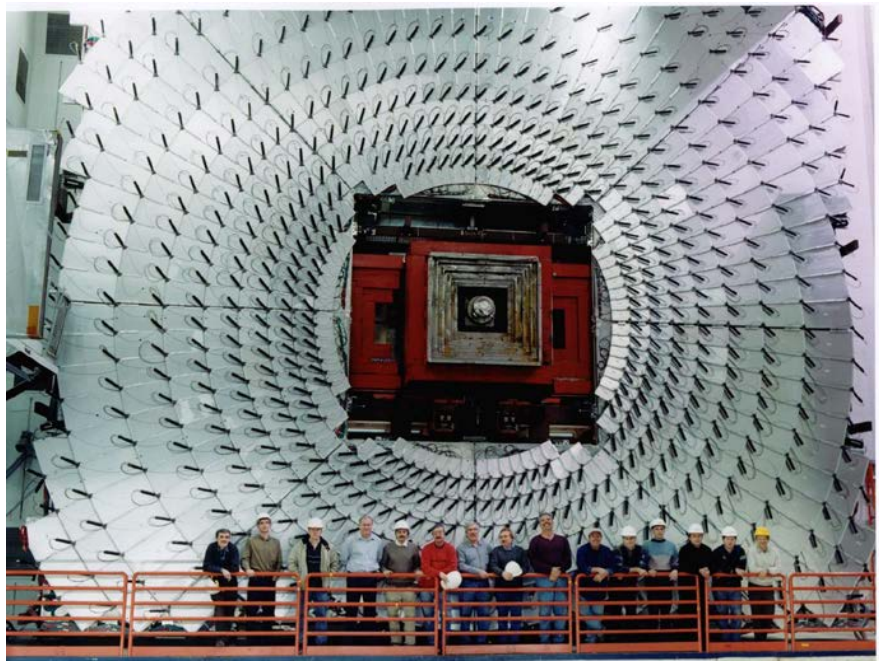

Fig. 2. C-layer of the forward muon trigger counters, the largest among the three layers, is approximately 12 x 10 m$^2$.

The use of WLS for light collection allows efficient light collection using PMTs with small photocathode diameter and provides possibility of better magnetic field shielding. The BICRON 404A scintillation light emission peak is 420 nm, its decay time is 2.0 ns, and its attenuation length is 1.7 m. The absorption peak of the Kumarin 30 WLS bar matches the emission peak of the scintillator. The light emission peak of the Kumarin 30 WLS bar is 480 nm, the decay time is 2.7 ns, and the attenuation length is 1.4 m.

Fig. 3 shows the details of the counter design. Two WLS bars, 4.2 mm thick and 12.7 mm wide, are placed along two edges of the scintillator plate with air gaps provided by narrow strips of adhesive tape attached to the scintillator plate as spacers. The end sections of both bars are bent by 44° to deliver light to the 25-mm-diameter FEU-115M phototube [7]. This phototube has an average quantum efficiency of 15% at 500 nm with a maximum gain around $10^6$. The PMT base consists of 13 resistors, first 3 resistors to dynodes close to photocathode are 2.4 MΩ, 1.8 MΩ and 1.3 MΩ, all others are 0.82 MΩ with anode load resistor of 1 kΩ. The opposite ends of the bars are made reflective using aluminized Mylar tape. To provide light tightness, the scintillator and WLS bars are wrapped in a layer of TYVEK [8] (type 1056D) and two layers of black photographic paper. The counter assembly is secured in a box made of two aluminum plates and an extruded aluminum profile along the perimeter. The number of photoelectrons for minimum ionizing varies from 240 for small A-layer counters to 70 for the largest C-layer counters. The uniformity of light collection over counter surface is ±10%.

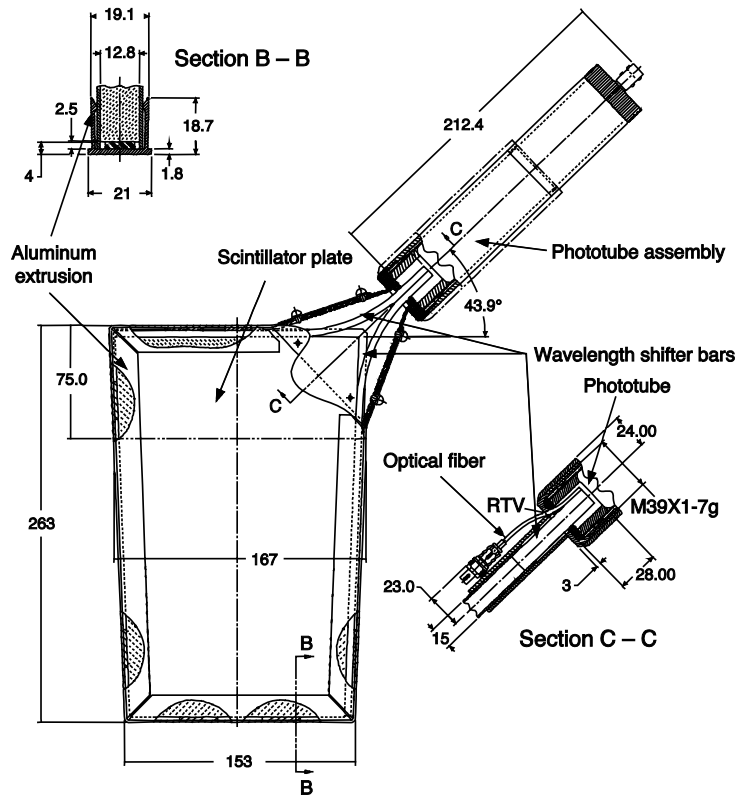

Fig. 3. Design of the forward muon scintillation counter. Dimensions are given in mm.

Effects of radiation damage on the light output of Kumarin 30 WLS bars and BICRON 404A scintillator were studied using prototype counters of different sizes [9]. The expected Run II highest radiation dose of 1.2 krad [4] near to the beam line of the A layer for luminosity of 12 fb$^{-1}$ (total luminosity delivered in Run II) is below what would degrade counters performance.

A different light collection design was used for approximately 150 counters located in confined areas near calorimeter cryogenic lines. These counters have twelve BCF 92 [10] multi clad WLS fibers along the edges of 12.8 mm thick BICRON 404A scintillator plates for collecting scintillation light [11]. Both ends of the fibers are glued into Plexiglas light-collector tubes, polished, and brought to a FEU-115M phototube located on top of the counter assembly. The same phototube assembly (Fig.3) and PMT base are used for these counters. This design provides signal of 60–110 photoelectrons per minimum ionizing particle, depending on the counter's size.

For each group of 16 counters the same high voltage power supply is used, so 288 power supplies modules [12] located in six VME crates were used for the 4214 pixel counters. For each power supply, the voltage and current values were monitored and recorded continuously during operation. Periodically, typically once a year, high voltage system was calibrated using high accuracy digital voltmeter. The operating voltage stability was ±1V, the difference between calibrations and readout values was ±7 V (full width). The current per phototube base was 145 µA at 1.85 kV and the total current per power supply was approximately 2 mA. Selection of photomultipliers for the counters has been performed to get close signal amplitudes for counters connected to the same high voltage power supply. The uniformity of $^{90}$Sr amplitudes for groups of 16 counters achieved using the matching procedure is ±25% (full width). High voltage values selected during the matching process are used for all Run II data taking with no adjustments. Some variations of phototubes gain in high voltage groups during years of running led to widening of the amplitude spread.

## 2. Counters and electronics failures

The performance of the scintillation counters was monitored on a regular basis during Tevatron runs. Voltages and currents of the phototubes high voltage power supplies were checked during every data taking shift using the D0 online monitoring system. Counters demonstrated stable operation and high reliability during physics data collection. About 20 counters have failed over the first 4 years of operation primarily due to bad high voltage or signal connectors and bases. Over the following years most of the failures were caused by poorly crimped LEMO connectors on 50 Ω coaxial signal cables. All cables were tested and some were fixed during shutdowns of the Tevatron. Over the last 4 years of operation, no PMT bases failed, and rare cases of non-working channels were caused by front-end electronic boards failures. Long accesses to the collision hall to repair failed counters occurred every 3–6 months. Movers designed to roll out fully assembled A- and B-layers of counters to provide access to all detector parts for repairs are mounted on the top of the layers. Typical duration of opening/closing of a counter layer including repairs is 1 day. Typical number of non-working counters during physics data collection was about 2 or 0.05%.

Compromised quality of the data during data collection was mainly caused not by the scintillation counters, rather by front-end electronic errors or by failures of low voltage power supply modules, located in the collision hall in front-end electronic VME crates. Typically we observed 4 cases of such failures per year in the forward muon scintillation system which contains 12 VME crates. Some (~30%, mostly in C-layers) of the power supplies failed at the beginning of the Tevatron stores during period of large beam losses and attributed to radiation damage, but most appear random. It took ~1 h access to the collision hall to replace a failed power supply, but if problem occurred at the beginning of a collider store, the data collected up to the access at the end of the store was compromised.

Compatible level of data losses was due to front-end VME crates errors caused by low level synchronization signal in VME crate. Typically, such errors cause a pause in date taking, around 0.5 min in duration, before an automated reset signal was issued.

## 3. LED calibration

LED-based calibration system was used to monitor the timing and gain of phototubes and electronics [13]. This system sends to PMTs light pulses similar to those of muon signals in amplitude and shape. Blue LED pulser modules distribute light signals to the PMTs via optical fibers, one fiber to each PMT. One pulser module per octant is used, 48 pulser modules in total for all 6 layers. To create a uniformly illumination, the light pulse generated by the LED goes through two stages of light mixing blocks. A PIN diode is mounted on the first light mixing block to provide an internal measure of the light intensity. The light pulses are split by the fiber block and fed to phototubes via clear plastic optical fibers with 1 mm diameter and ~4 meters long. An optical connector is mounted on each counter for connection of the optical fiber from pulser module to the piece of the fiber directed to the PMT photocathode. The LED calibration system has fast response, served well during commissioning of the counters and was widely used for monitoring of PMTs gains and electronics between Tevatron stores.

Full LED calibration for all pixel counters was performed about once a year. Such calibration took place in May of 2001 for the first time, and these results became the reference set for all the subsequent calibrations. The procedure used for the LED data analysis is the following: for each channel the pedestal value is subtracted from the LED amplitude value (the same subtraction was made for the 2001 reference set), then the resulting amplitude for the current LED calibration was divided by the one of the reference set and then this ratio is normalized by the ratio of PIN diode amplitudes. Obtained distributions are fitted by Gaussian to obtain mean value and width. For the calibration of timing, for each channel the fitted mean value of time of the reference 2001 calibration is subtracted from the fitted mean time of the current calibration.

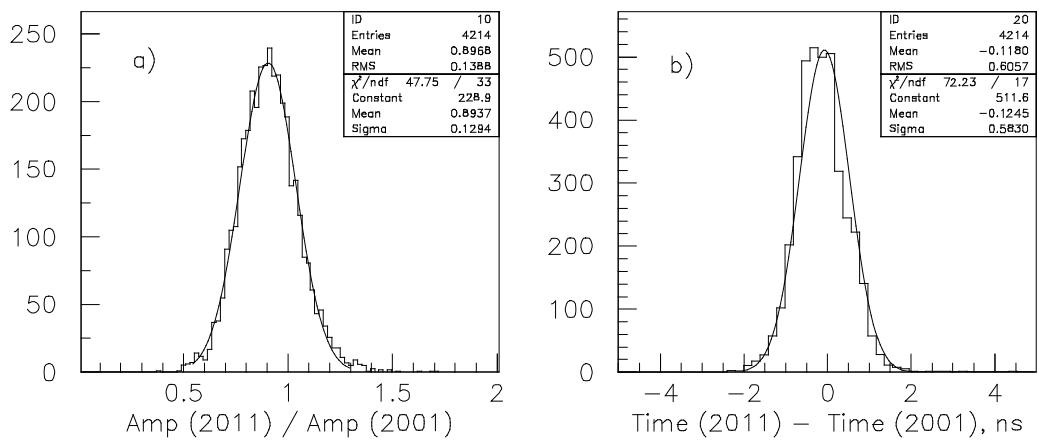

Fig. 4. 2011 LED calibration results for scintillation counters:
(a) amplitude ratio distribution and (b) time difference distribution.

The results of the LED calibration made in the year 2011, at the end of the Tevatron run, are shown in Fig. 4. Fig. 4a is the amplitude ratio distribution for all counters, Fig. 4b is the time difference distribution. Mean value for the amplitudes ratio is 0.90 and sigma of the Gaussian fit is 0.13. Mean time difference is –0.12 ns with sigma of 0.6 ns. Figs. 5 and 6 demonstrate LED calibration results over a period of ten years: average amplitude ratio and average time difference. The first point on each plot is the year 2001 reference point. The amplitude ratio is slowly decreasing with time and total delivered luminosity with the reduction reaching 11% in year 2011. Gaussian standard deviation for the amplitude ratio is slowly increasing in time. The accuracy of the normalization procedure for mean values of amplitude spectra is better then 1% because of large statistics used for both calibrations. The variations in phototube gains during years of running led to widening of the amplitude spread and increased the width (sigma) of the amplitudes ratio to 13% in year 2011. The decrease of averaged amplitude by 11% and the increase of Gaussian sigma to 13% mean that the amplitude for single scintillation counter changed over 10 years by –11±13%. The LED calibration time variations do not exceed 0.26 ns with sigma of 0.6 ns. Maximum timing shift of 0.26 ns was after 1.5 years and mainly due to LED drive boards and scintillation front-end electronic replacements.

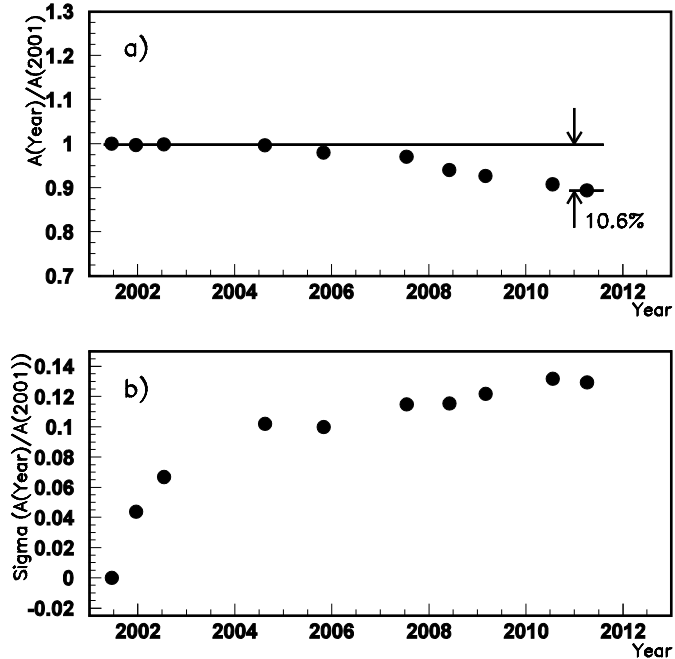
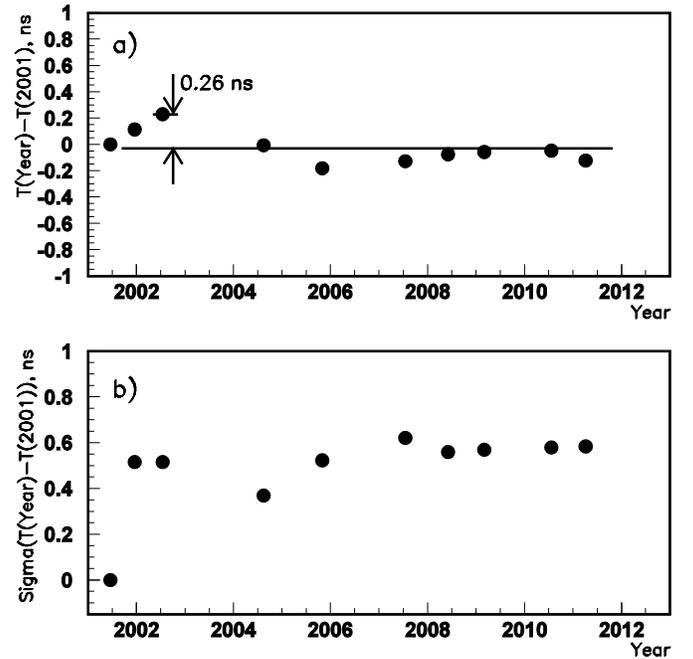

Fig. 5. Scintillation counters LED calibration amplitude distribution parameters over ten years. The line shows constant ratio 1. Amplitude reduction is ~11%. Statistical errors are less than the size of the data points.

Fig. 6. Scintillation counters LED calibration timing distribution parameters over ten years. 0.26 ns is the maximum deviation from the straight line is constant fit. Statistical errors are less than the size of the data points.

Fig. 7 shows forward muon counters LED calibration amplitude parameters up to March 2011 (10.5 fb$^{-1}$ of total delivered luminosity) versus total integrated Tevatron luminosity [14] for all counters and for A, B and C layers separately. The data point at 0.6 fb$^{-1}$ is shown on this figure for all counters only, it was not processed for layers separately due to low, ~1% amplitude changes.

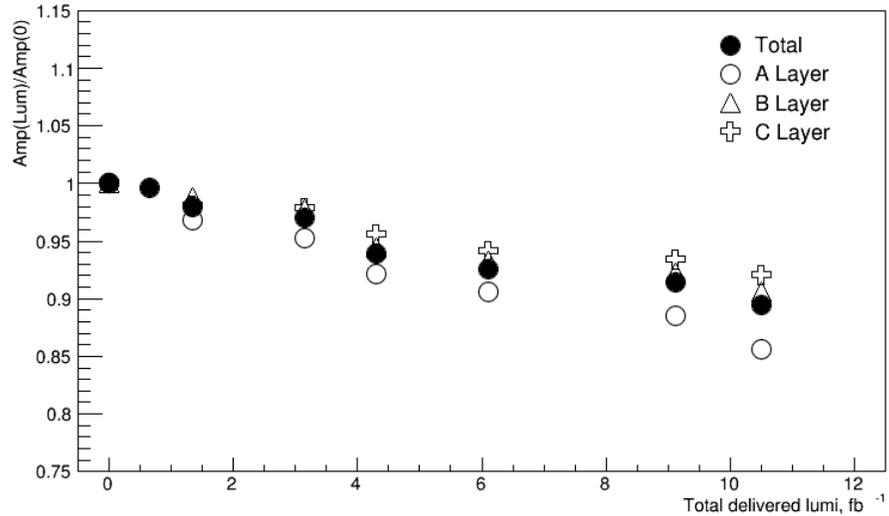

Fig. 7. Scintillation counters LED calibration amplitude dependence versus total Tevatron integrated luminosity for all counters and for A, B and C layers separately. Statistical errors are less than the size of the data points.

Differences in the amplitude decrease from layer to layer are significant and can not be explained by the measurement uncertainties. For the last calibration shown in this figure, amplitude ratio is 0.91 and 0.92 for B- and C-layers and 0.86 for A-layer counters. This result will be discussed later in comparison with amplitude stability measurements using muons originating from proton–antiproton interactions.

## 4. Amplitude and time measurements of muons from proton–antiproton collisions

Amplitude measurements for muons originating in proton–antiproton collisions were also used for monitoring the stability of the counters operation. Front-end electronics ADCs are multiplexed so single amplitude measurement in a group of 16 counters is performed for each triggered event. Large statistics is needed for amplitude measurements because of multiplexed ADCs and relatively low occupancy of the scintillation counters. This calibration tests not only PMTs and electronics as LED calibration but also the wave-length shifter and the scintillator of the counter.

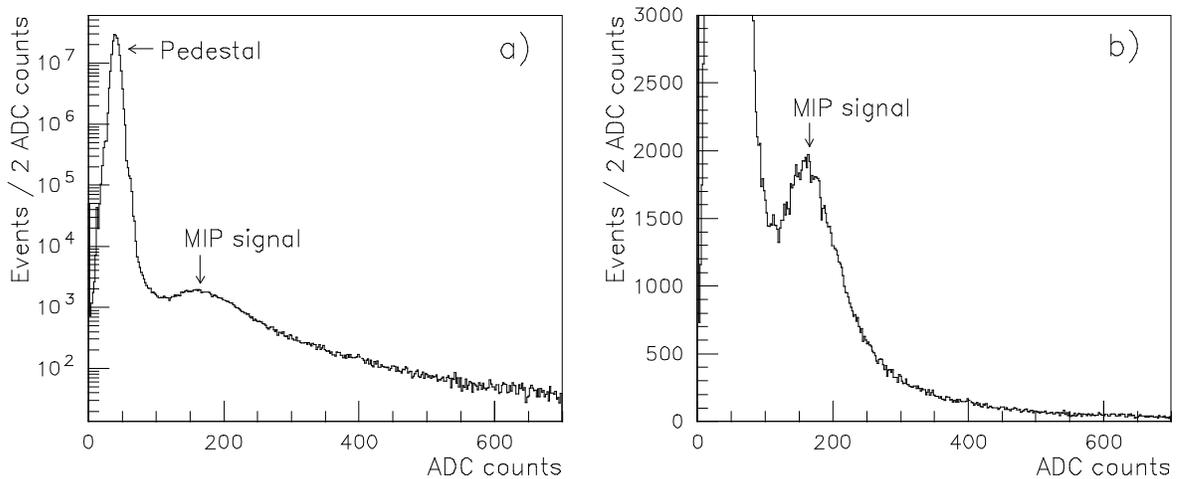

Fig. 8. Raw amplitude distributions for all 4214 counters (December 2002).

Signal amplitudes from the scintillation counters are measured during the experiment physics data taking. Muon on-line monitoring program is used and all muon triggers are used for the event selection. Raw amplitude distributions are presented in Fig. 8. Pedestals are equal to 40 ADC counts (Fig. 8a) and in an expanded view (Fig. 8b) the peak from minimum ionizing particles (MIP) is visible at ~165 ADC counts. A procedure which uses correlations between hits in counters in 3 layers was developed to select hits from muons: in an event hits in each of A, B and C layers in the neighboring counters are required: ±2 counters in η direction and ±1 counter in φ. No timing cuts are used for this selection. This simple procedure of "reconstruction" provides clear MIP peak due to the low occupancy (discussed below) of the counters.

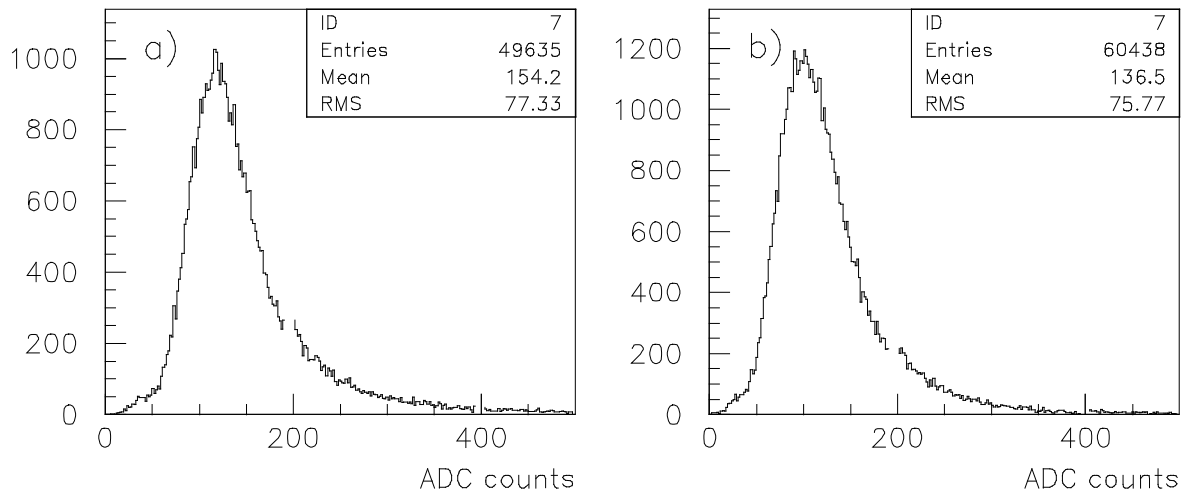

Fig. 9. "Reconstructed" muons amplitude distributions for all 4214 counters for (a) December 2002 (b) March 2011. Pedestals are subtracted.

Muon amplitudes measurements using procedure described above was performed first time in December 2002. Final check of the stability of the counters response with the same procedure was performed in March 2011. Fig. 9 shows amplitude distribution for all 4214 counters for December 2002 and March 2011 measurements. Pedestals are subtracted. It was shown that >98% of "reconstructed" by the described procedure muons have amplitudes corresponding to MIP energy deposition. The electronics discriminator threshold is equal to 7 mV or about 25 ADC channels. Fig. 9 shows good uniformity in the signal amplitudes as well as a reasonable choice of discriminator thresholds. To monitor the stability of the scintillation counters calibration procedure was performed periodically, typically once a year. Scintillation counters amplitude ratio Amp(Year)/Amp(2002) over 8.5 years is shown in Fig. 10, the mean values of amplitude distributions are used. Statistical errors are less than the size of the data points. The averaged amplitude decrease for North and South layers is the same within 1%. Amplitude decrease is ~1.8% per year over last 7 years period.

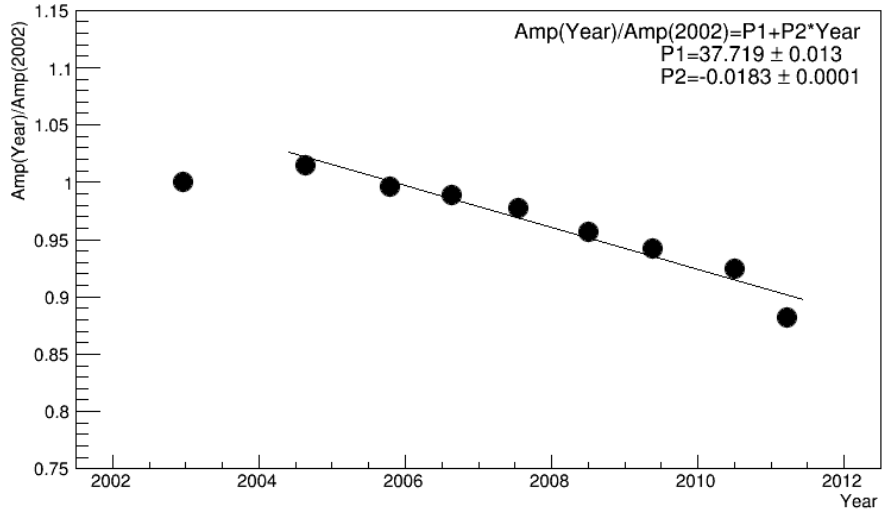

Fig. 10. Forward muon counters amplitude ratio Amp(Year)/Amp(2002) over 8.5 years. P1 and P2 are the straight line fit parameters. Statistical errors are less than the size of points. Amplitude decrease is 1.8% per year for the last 7 years period.

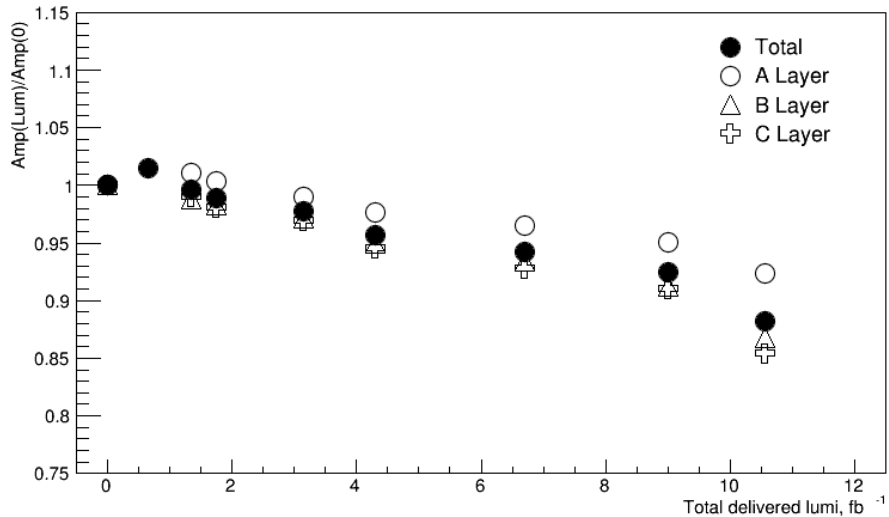

Fig. 11. Scintillation counters amplitude ratio Amp(Lum)/Amp(0) variations versus Tevatron luminosity for A, B and C layers separately and averaged together. The data point at 0.6 fb$^{-1}$ is shown on this figure for all counters only as it was not processed for layers separately. Statistical errors are less than the size of the data points. North and South side layers show similar results.

Fig. 11 shows scintillation counters mean values of amplitude changes obtained using muons from collisions versus total delivered luminosity for all layers averaged and for A, B and C-layers separately. The variations of the ratio Amp(Lum)/Amp(0) for both LED calibrations and for muons originating from proton–antiproton collisions are presented versus delivered luminosity in Fig. 12. Presenting results versus total delivered luminosity shows linear dependence and is an indication that decrease in amplitude is due to the radiation aging of the scintillation counters. The average value of amplitude ratios for all 4214 counters of all layers has decreased by 11% over a period of ten years. LED and MIP measurement results are in reasonable agreement with each other and both show rate of aging about 1% per year.

While it is important to determine which component of the scintillation counter is responsible for aging, we do not have direct measurements of stability for each component for decisive conclusion. The analysis of differences in aging rates for A, B and C layers using Fig. 7 and Fig. 11 and differences in counters sizes and different radiation conditions for the layers are used to identify main aging components. LED monitoring light pulses directly illuminate PMT photocathode, so the decrease of amplitudes may be explained by aging of PMT, by the decrease of the attenuation length of ~ 4 m long fiber delivering LED pulses to PMT, or by imperfect normalization of LED amplitude to PIN-diode signal. Muon calibration tests all parts of the scintillation counter and sensitive to scintillator, WLS bars and PMT aging. Muon calibration is expected to be more relevant to the counter properties than LED monitoring.

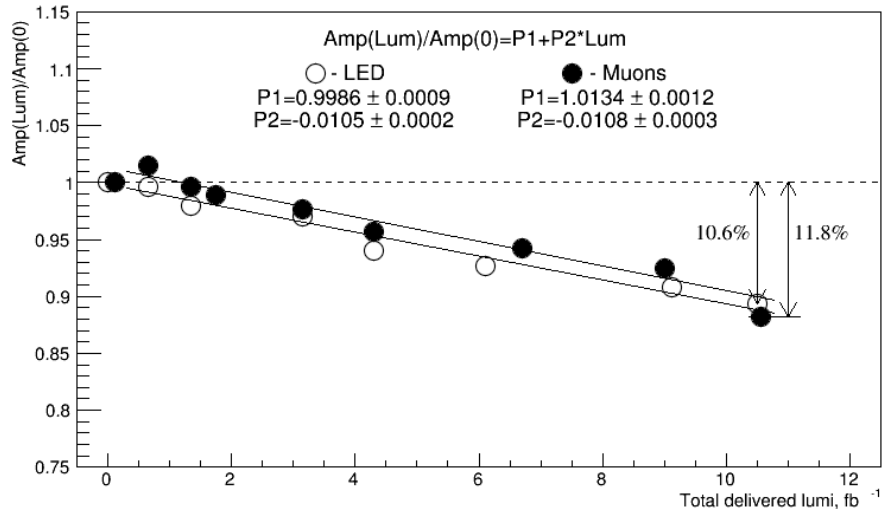

Fig. 12. Forward muon scintillation counters amplitude ratio Amp(Lum)/Amp(0) versus total delivered luminosity for LED and muon calibrations. P1 and P2 are the straight line fit parameters. Statistical errors are less than the size of the data points.

Averaged for all layers results in Fig. 12 demonstrate similar amplitude drop for both LED and muon calibrations. PMT gain drop is the only common part for LED and muon calibrations and could be considered as the main reason for the amplitude decrease. But detailed comparisons using differences in aging for A, B and C layers including different counter sizes show that situation is more complicated. The amplitude drop for LED calibration as shown in Fig. 7 is 14%, 9% and 8% for A, B and C–layers respectively. The results of the calibration using muons as shown in Fig. 11 have opposite behavior with amplitude decrease 8%, 13% and 15% for layers A, B, and C, respectively. As a result PMT gain loss only could not explain observed results, at least two components should be considered to explain the observed behavior of the LED and muon calibrations.

The PMT aging is sensitive to the anode current and to the accumulated electric charge on the anode. Signal amplitudes are similar for all PMTs, so PMT aging depends on counting rates. Counting rates for A-layer counters are 2–3 times higher than for B- and C-layers as they are proportional to counters occupancy shown later in Fig. 16. Faster PMT aging is expected for A-layer in comparison to B and C-layers in agreement with results using LED calibration and contrary to the results using muons. This consideration excludes PMT aging as the only reason of the observed amplitude drop as well.

WLS bars aging due to the radiation damage is expected to give less amplitude drop for A-layer counters due to the shorter length of bars. The counter (and WLS bars) sizes ratio is 1:1.6:1.9 for A, B and C-layers and the size of the largest counter in C-layer is 60x110 cm$^2$. Assuming no PMT aging, aging of WLS bars (or scintillator) could explain muon calibration results. But ageing of WLS and scintillator could not explain decrease observed in LED calibrations.

Attenuation length decrease (due to radiation damage) of fiber optic lines of LED pulser system could explain LED calibration results differences for A, B and C layers. For A-layers radiation rate is higher and leads to faster degradation versus delivered luminosity. Above considerations provide options for the explanation of the observed amplitude decrease in both muon and LED calibrations, but actual cause or causes cannot be more definitively identified based on the available information. ~10% decrease in the scintillation counters amplitudes after 10 years of operation has no effect on the physics performance of the D0 forward muon system.

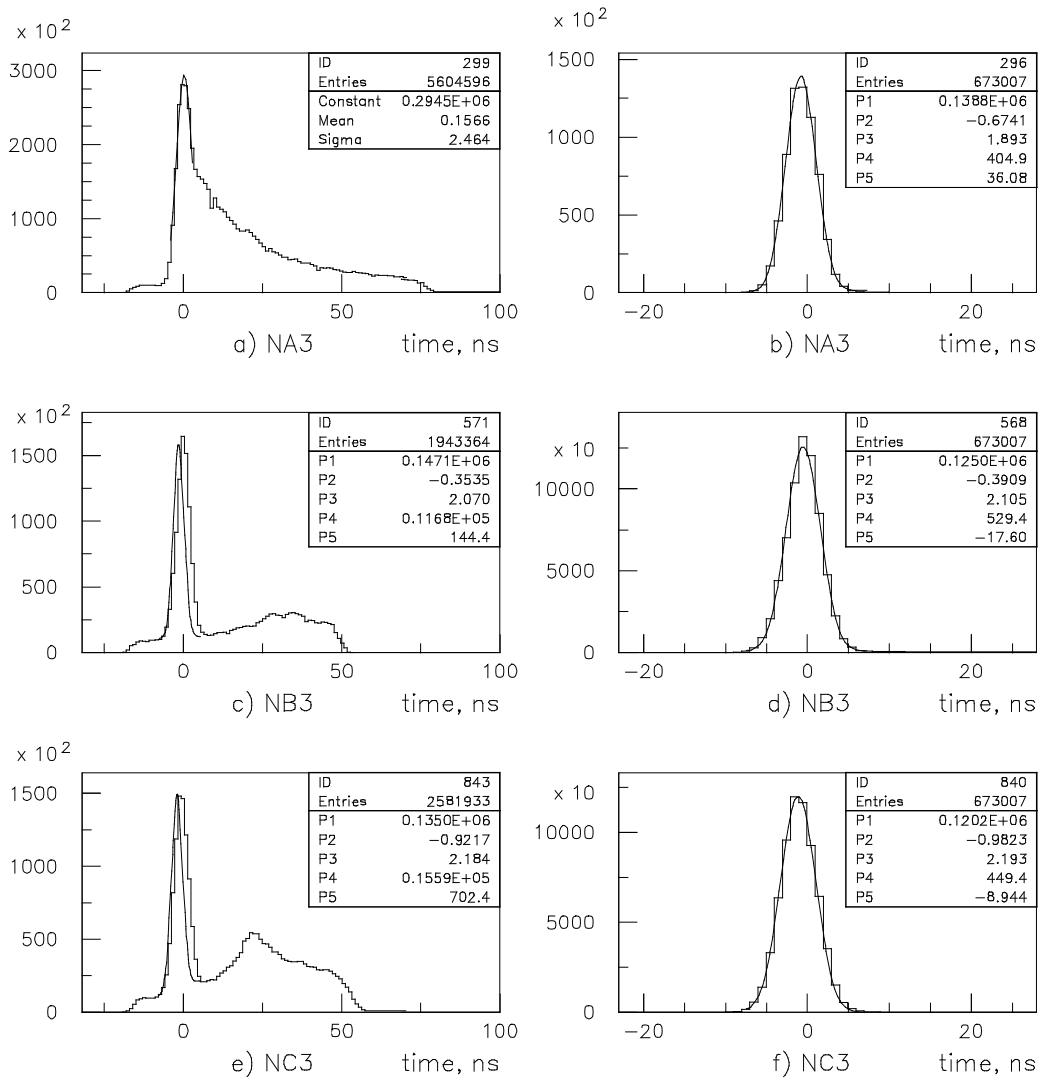

Fig. 13. Raw (left) and "reconstructed" muons time distributions (right) for octants North A3 (a,b); North B3 (c,d); North C3(e,f) for August 2008 measurement. The parameters P1-P3 are Gaussian fit parameters (P1: normalization, P2: mean value, P3: sigma) and P4 and P5: linear fit parameters of the background.

Timing properties of the counters were also studied using "reconstructed" muons. Fig. 13 presents an example of the measurements made in August 2008. Raw (for all recorded events) and "reconstructed" muon time distributions are shown for one of the octants of each North layer. The mean and sigma parameters of a Gaussian fit for B- and C-layers are consistent for raw and "reconstructed" events, but for A-layer fit parameters for raw events are distorted by high level of background. The time resolution (sigma) parameters are 1.9, 2.1 and 2.2 ns for A, B and C-layer octants. The difference is caused by different counters sizes in A, B and C layers as time measurement does not take into account location where muon passed through the counter, so smaller counters have better time resolution. The time resolution was measured during counters production and found to be from 0.5 to 0.9 ns depending on the counter size. The timing accuracy is determined mainly by the spread of PMT delays and signal cable lengths as well as by scintillator front-end electronics with bin width of 1 ns. Monitoring of the stability of timing measurements and position of the peak using muons from collisions is limited by seasonal drift of the D0 clock signal as described below. For short periods of time, about one hour, muon timing peak position found from Gaussian fits has stability of better than 0.05 ns.

## 5. Beta-source calibration results

Calibration using $Sr^{90}$ source is useful and efficient tool for direct and fast monitoring of counters amplitudes. During commissioning of the forward muon scintillation counters system in 1999–2000 all counters in each of 48 octants were matched in 6 groups with single HV power supply for each group. Matching included measurements of PMT pulse amplitude using $Sr^{90}$ beta source. The maximum value of amplitude signal, the high end of the beta spectrum, was measured for each counter using oscilloscope. The accuracy of the measurement is ~ 10%. Results obtained in 1999–2000 are used as a reference set for all the subsequent $Sr^{90}$ calibrations. Periodically, once a year, during long accesses to the collision hall $Sr^{90}$ source measurements were repeated in order to check the stability of the counters operation for a sub-set of counters. One of such source calibrations, performed in May 2006, shows ~4% decrease of average amplitude for studied counters in comparison with measurements during matching, which is within measurement uncertainties in agreement with results presented above.

No systematic drift in pixels amplitudes for $Sr^{90}$ calibration between 1999 and 2009 measurements is observed within 10% measurement accuracy.

## 6. Long-term stability of muon timing peak and D0 clock adjustments

Scintillation counters system provides muon triggering as well as time of flight measurement with ~1 ns accuracy. Stability of muon time measurement is important for muon identification, background rejection, and new particles searches [15, 16]. To perform muon timing analysis we fit with Gaussian function time peak of muon hits detected in each layer of the muon system, in each octant, high voltage group (16 counters) or even for each individual counter for detailed studies. In Fig. 13 time distributions for all hits detected in three selected octants are presented. The time distributions for B and C-layers both have narrow peaks and are more useful for the analysis of peak position stability, A-layer has higher background and muon peak is less pronounced. Timing constants T0 were loaded to scintillation counters front-end boards and subtracted on-line from measured times to keep peak position for relativistic muons originating in proton–antiproton collisions at "zero time". T0 value is defined for each group of 16 counters connected to the same HV source to compensate differences in signal cables lengths and times of flight. Raw time distributions for all B-layers (both North and South) and C-layers (both North and South) counters are shown in Fig. 14 including Gaussian fit results for the peaks.

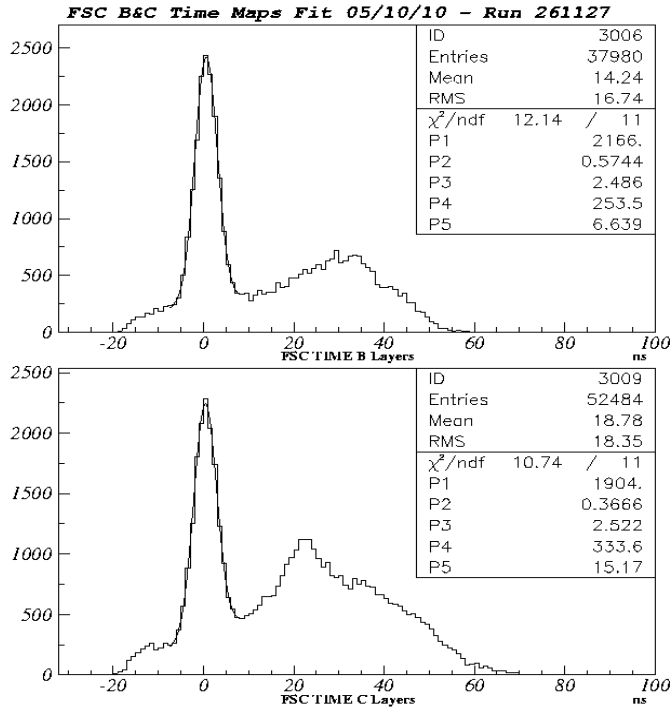

Fig. 14. Raw time distributions for all B and C-layer counters. P1–P3 are Gaussian fit parameters, P2 is the mean value of the time peak.

Variations in the muon timing peak obtained from Gaussian fits were observed at the beginning of Run II with peaks drifting from zero over a period of ~2 months. New T0 values were loaded to keep muon timing peaks at zero several times during first 1.5 years of Run II. Studies of the D0 clock drift and correlation studies between the clock drift and forward muon scintillation counters timing were performed. It was found that variations are mainly due to the D0 clock drift with respect to the proton–antiproton collisions.

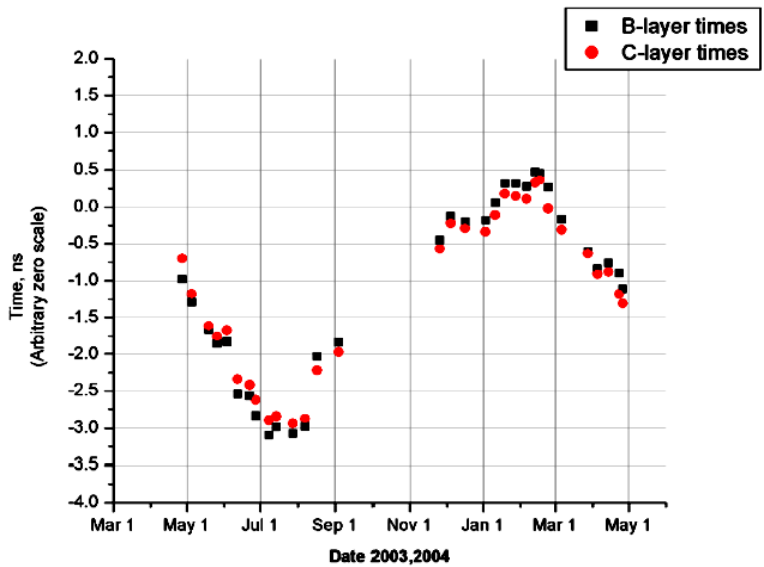

Figure 15. Muon timing peak variation over one year with no T0 adjustments and no clock adjustments

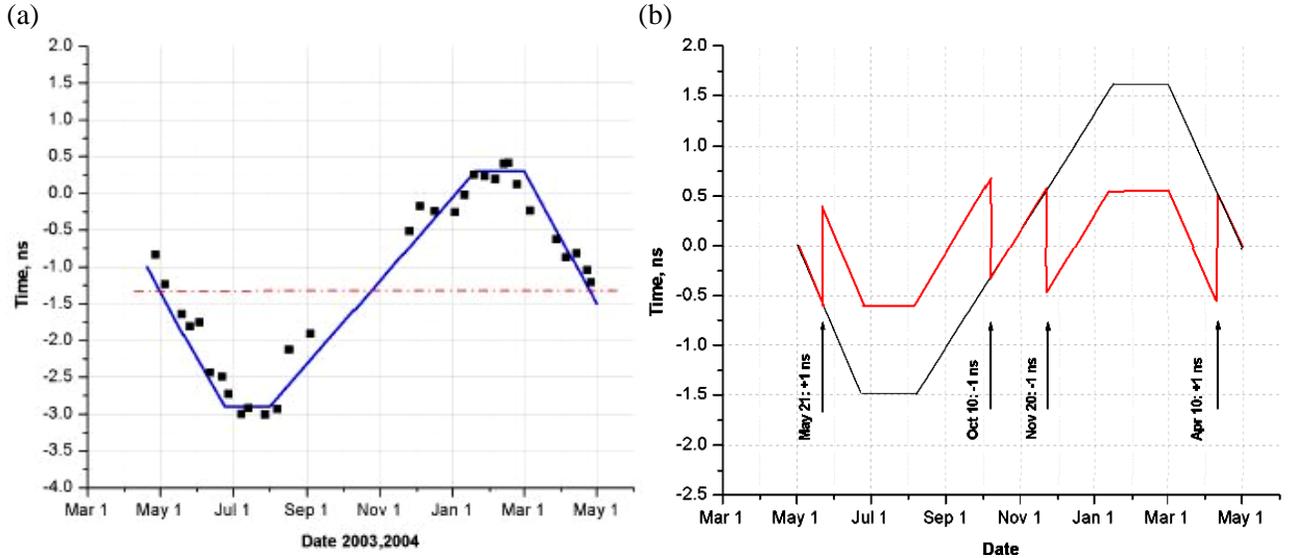

Fig. 16. D0 clock adjustment. (a) approximation of muon timing drift, (b) scheme of the clock seasonal adjustments.

The main source of timing instability is seasonal drift of the D0 clock. Figure 15 shows muon timing peak (Gaussian fit) position variation over one year for averaged values for all B and C-layers counters. This figure shows timing drift as it would look like with no T0 adjustments and no D0 clock adjustments. With no timing corrections variations in seasonal drift of muon timing peak reach 3.6 ns. The timing drift can be explained by seasonal temperature changes. The clock signal is transmitted via ~2 km long quartz fiber. This fiber is placed ~2 feet underground between the accelerator clock generating facility and the experiment and was therefore sensitive to outside temperature. There are "flat" periods during summer and winter when temperature is stable and there are steep slopes during spring and fall when temperature rapidly changes.

Clock adjustment is easy to implement and corrects timing not only for the muon system, but for the rest of the experiment as well. The scheme of the adjustments described below minimizes timing drift and the number of the adjustments. The timing variations in Fig. 15 can be approximated by a set of straight lines presented in Fig. 16a. In April of 2004 D0 clock adjustment was made by change of the cable delay to move "zero-line" of Fig.16a from –1.3 ns close to 0 ns. After that ~4 adjustments per year have been made: around May 20, October 10, November 20 and April 10, see Fig. 16b. This scheme of the D0 clock adjustments kept the muon time variations within $\Delta t = \pm 0.6$ ns. Real adjustments dates were weather dependent and chosen by continuously monitoring timing peaks for B and C-layers.

## 7. Counters occupancy

The occupancy of the detector channels defined as the average number of hits per event per channel, which depends on type of an event and triggers used, is important for estimates of the detector efficiency losses and radiation aging rate. High level of detector occupancy may limit in some cases the luminosity for physics data collection and could cause detector parameters deterioration.

The shielding assemblies shown in Fig.1 consisting of layers of iron, polyethylene, and lead were installed during D0 Run II upgrade. As a result, the number of background hits in the muon counters was reduced by a factor of between 40 and 100 in comparison with the absence of the shielding [4].

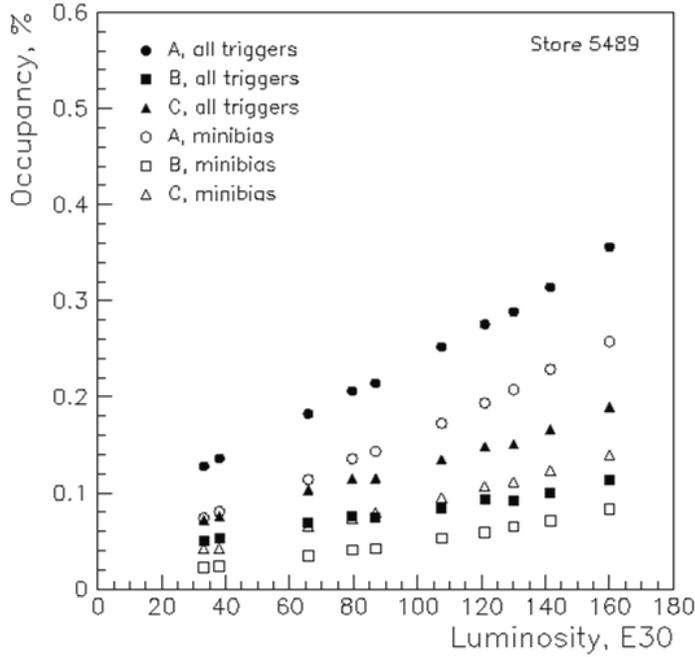

Fig. 17. Forward muon scintillation counters occupancy for minimum bias triggers and for all triggers in physics data collection runs for A, B and C–layers.

D0 forward muon scintillation counters occupancy for minimum bias triggers and for all triggers in physics data collection runs for A, B and C–layers are shown in Fig. 17. Minimum bias trigger requires evidence of proton–antiproton collision and reflects properties of the majority of collisions. The occupancy for minimum bias trigger is useful for average counting rate and radiation doses estimates. The dependence of occupancies from luminosity for minimum bias trigger is not extrapolating to zero at low luminosity because events are selected when at least one collision happened per beam crossing.

The occupancy of counters is low in accordance with the expectations based on shielding design. For Tevatron Run II the bunch crossings were separated by 396 ns, with some empty spaces for beam injection, the collision rate was $1.73 \cdot 10^6$ collisions per second and the minimum bias occupancy for A-layer was 0.26% at luminosity $160 \cdot 10^{30}$ cm$^{-2}$s$^{-1}$. Under these conditions, the average counting rate for A-layer counter was $1.73 \cdot 10^6$ s$^{-1} \cdot 0.263 \cdot 10^{-2}$=4.4 kHz. Even for highest Run II luminosity of $43 \cdot 10^{32}$ cm$^{-2}$s$^{-1}$, the average counting rate did not exceed 11 kHz. The calculated value of the average PMT anode current for the above conditions is 0.1 µA for signal amplitude 30 mV and 15 ns pulse duration. The counting rate of 11 kHz at luminosity $43 \cdot 10^{32}$ cm$^{-2}$s$^{-1}$ corresponds to the average charge accumulated on PMT anode of 3 C and average radiation dose of 0.03 krad for typical A-layer counter (due to MIP particles) for the delivered luminosity of 12 fb$^{-1}$. The radiation dose for the closest to the beam counters from MIP is estimated at 0.2 krad. Both accumulated charge and radiation dose estimates are below expected to degrade counters performance.

The occupancy could be higher due to event selection. For muon triggers used for forward muon special runs, the occupancy was two times higher than for minimum bias trigger, but did not exceed 1% for highest Run II luminosity.

## 8. Monitoring of the performance of the forward muon system using single muon yields

Single muon yield measurements were performed on a regular basis for monitoring the stability of the entire forward muon system (scintillation counters and mini-drift tubes) as well as the stability of muon trigger and the efficiency of the reconstruction program. Single muons yield is defined as number of reconstructed muon tracks normalized to the integrated luminosity for this data set. Periodically, typically once a month and a half, special data was collected with ~100-150 thousands events using single muon triggers. This trigger selects single muon events using information from the muon system only, other detector systems are not included. These events are further processed using the off-line reconstruction program [17]. For yield calculation, muons reconstructed using muon detectors only are selected.

Yield results for runs collected between July 2006 and September 2011 are presented in Fig.18 demonstrating operation of the forward muon system for the period of 5 years with stability of 1%. Uncertainty on the yield $\sigma_Y$ is set to 1% if statistical error is less than 1%, and equal to statistical error, if statistical error is above 1%. This definition covers systematic uncertainties of approximately 1%.

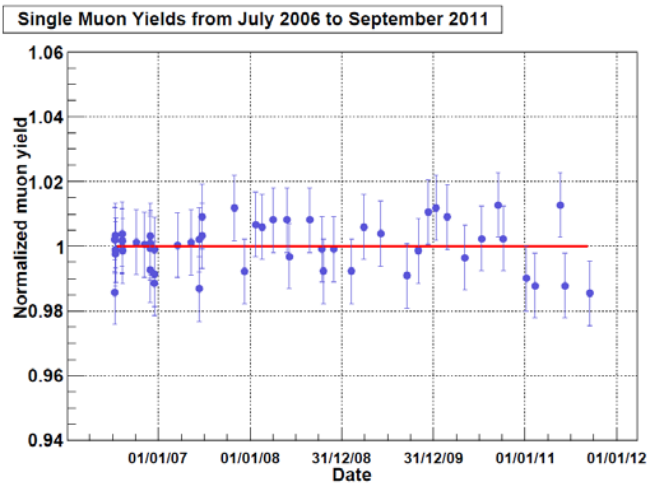
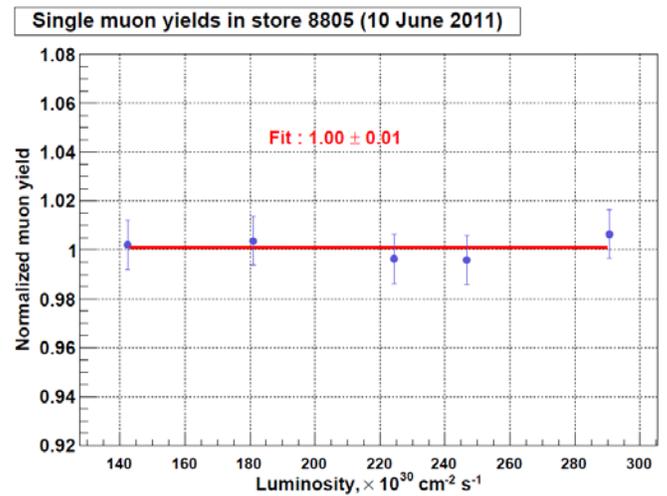

Fig. 18. Normalized muon yields versus time.    Fig. 19. Muon yields versus luminosity.

Forward muon special runs have been taken to study the luminosity dependence of the forward muon yields. Result from special runs recorded on June 10, 2011 is shown in Fig.19. No muon yield dependence on instantaneous luminosity is found at the 1% level for this and other similar studies for the luminosity range (20–300) $310^{30}$ cm$^{-2}$s$^{-1}$.

## 9. Summary

The performance of the D0 Run II forward muon scintillation counters is presented. The counters system was monitored regularly during collider Run II and demonstrated stable operation and high reliability during 10 years of physics data collection.

An LED based calibration system was used to monitor timing and gain of phototubes and front-end electronic tests. Results of LED calibrations made from 2001 to 2011 are presented. The signal amplitude ratio average value decreased over a period of 10 years by 12%.

Procedures for counters amplitude and timing calibrations using muons originated from proton–antiproton collisions are described. Results of counters amplitude monitoring for the years 2002 to 2011 are presented. The average value of counter amplitude for muons has decreased by 11% over a period of 10 years. Radiation aging was low and did not affect substantially counters performance.

Counters timing drift was found to be strongly correlated with the D0 clock signal drift which is correlated with seasonal temperature changes. Adjustments made four times a year using muon timing peak provided the stability of the D0 clock system within ±0.6 ns.

The occupancy of the scintillation counters was low during physics data collection. For the highest Run II luminosity of $4 \cdot 10^{32}$ cm$^{-2}$s$^{-1}$, the occupancy of counters is less than 1%. Monitoring of single muon yields demonstrates stable operation of the forward muon system in Run IIb. No dependence of muon yield from luminosity is observed for the luminosity range $(20-300) \cdot 10^{30}$ cm$^{-2}$s$^{-1}$.

## Acknowledgements

We thank our D0 colleagues and staffs at Fermilab and collaborating institutions for their support with construction of the forward muon system, collection of data and data processing.

## References

[1] A. Artikov *et al.*, Nuclear Instruments and Methods A 672 (2012) 46.
A. Artikov *et al.*, Nuclear Instruments and Methods A 579 (2007) 1122.
[2] A. Artikov, D. Chokheli, Nuclear Instruments and Methods A 591 (2008) 468.
[3] V. Abramov *et al.*, Technical Design Report for the DØ forward trigger scintillation counters, Nuclear Instruments and Methods A 419 (1998) 660.
[4] V.M. Abazov *et al.*, Nuclear Instruments and Methods A 552 (2005) 372.
[5] Bicron Corporation, 12345 Kinsman Rd, Newbury, OH 44065-9677, USA.
[6] S. Belikov, et al., Physical characteristics of polymethylmethacrylate scintillator SOFZ-105, Preprint IHEP 92-55, Instruments and Experimental Techniques 36 (1993) 390.
[7] MELZ FEU, Ltd.,124460, Moscow, Zelenograd, 4922 st, bld. 4/5, Technopark "ELMA-PARK". http://www.melz-feu.ru/products/?id=50
[8] Dupont De Nemours & Co., 705 Canter Rd., Rt. 141, Wilmington, DE 19810-1025, USA.
[9] V. Bezzubov *et al.*, Fast scintillation counters with WLS bars, Proceeding of SCIFI 97 conference, p.210, South Bend, Indiana, 1997. AIP Conf. Proc. vol. no 450, (1998) p. 210.
[10] Saint-Gobain Crystals, 17900 Great Lakes Parkway, Hiram, OH 44234
[11] V. Evdokimov, Light collection from scintillation counters using WLS fibers and bars, Proceeding of SCIFI 97 conference, p.300, South Bend, Indiana, 1997. AIP Conf. Proc. vol. no 450, 1998, p. 300.
[12] S.-C. Ahn, et al., A new VME based high voltage power supply for large experiments In: Proceeding of Conference Record of the 1991 IEEE Nuclear Science Symposium and Medical Imaging Conference, vol. 2, (1991) p. 984.
M.-J Yang, The DØ high voltage system, Nuclear Physics B - Proceedings Supplements, vol. 23(1) (1991) 402.
[13] P. Hanlet *et al.*, Nuclear Instruments and Methods A 521 p. 343 (2004).
[14] B.C.K. Casey *et al.*, , Nuclear Instruments and Methods A **698**, p.208 (2013).
[15] V. M. Abazov *et al.* Physics Review D 87 (2013) 052011.
[16] V. M. Abazov *et al.*), Physics Review Letter **102**, (2009) 161802.
[17] V. M. Abazov *et al.* Nuclear Instruments and Methods A **737** (2014) 281.